\documentclass [prb,twocolumn,amsmath,amssymb,longbibliography]{revtex4-2}
\setcitestyle{numbers,square}
\usepackage{appendix}
\usepackage{array}
\usepackage{amsmath,amssymb}
\usepackage{tabularx}
\usepackage{graphicx}
\usepackage{bmpsize}
\usepackage{bm}
\usepackage{color}
\usepackage{amssymb}
\usepackage [version=3]{mhchem}
\usepackage{newtxmath}
\usepackage{hyperref}
\usepackage{url}
\usepackage{physics}
\usepackage{here}
\usepackage{float}
\usepackage{comment}

\DeclareMathAlphabet{\mathpzc}{OT1}{pzc}{m}{it}

\def\para{\parallel}

\def\vv{\vb*}

\newcommand{\mage} [1]{\textcolor{magenta}{ #1}}

\newcommand{\rev} [1]{\textcolor{black}{ #1}}

\begin{document}

\title{
{
Twist-angle tunable Josephson junctions in three-dimensional superconductors}
}

\author{Tenta Tani}
\affiliation{Department of Physics, The University of Osaka, Toyonaka, Osaka 560-0043, Japan}
\author{Takuto Kawakami}
\affiliation{Department of Physics, The University of Osaka, Toyonaka, Osaka 560-0043, Japan}
\author{Mikito Koshino}
\affiliation{Department of Physics, The University of Osaka, Toyonaka, Osaka 560-0043, Japan}

\date{\today}

\begin{abstract}
{
We theoretically investigate the superconducting phase and perpendicular Josephson supercurrent in twisted three-dimensional (3D) superconductors, where two layered 3D materials are stacked with a relative twist.
We formulate the Bogoliubov–de Gennes Hamiltonian and develop a self-consistent method to calculate the superconducting order parameter and the resulting supercurrent.
Applying this framework to a toy model with Fermi surfaces located near the Brillouin zone corners, we demonstrate a phase discontinuity at the twisted interface, indicating that a Josephson junction is formed purely by the twist.
Our calculations reveal that the interface supports a finite critical current even when the Fermi surfaces of the two superconductors are completely separated, unlike in the case of a twisted normal-metal interface.
We further show that the critical current can be effectively controlled by the twist angle, transitioning from a high-transparency regime at small angles to a low-transparency regime at larger angles.
}
\end{abstract}


\maketitle
\section{Introduction}
\label{sec:intro}

\begin{figure}
\begin{center}
   \includegraphics [width=0.5\linewidth]{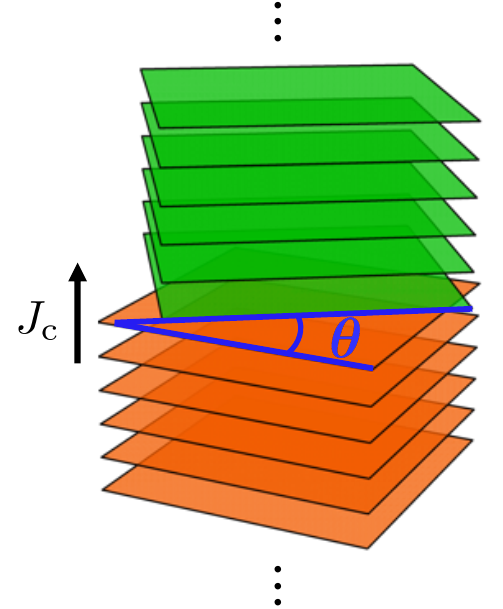}
   \caption{
   Schematic illustration of a twisted three-dimensional superconductor with a twist angle $\theta$.
   The black arrow indicates a possible Josephson supercurrent flowing in the perpendicular direction.
            }
\label{fig:schematic-3dsystem}
\end{center}
\end{figure}

Twisted two-dimensional (2D) materials have attracted considerable attention in the field of condensed matter physics.
Typical examples include twisted bilayer graphene (TBG)~\cite{Bistritzer2011,Cao2018SC,Cao2018Ins,Yankowitz2019,sharpe2019emergent,serlin2020intrinsic,stepanov2021competing,Kerelsky2019maximized,Xie2019spectroscopic,Jiang2019charge,Choi2019electronic,Polshyn2019large,Lu2019super,Cao2020strange,Saito2020independent,Zondiner2020cascade,Wong2020cascade,Arora2020super,Stepanov2020untying} and twisted transition metal dichalcogenides~\cite{liu2014evolution,van2014tailoring,yu2015anomalous,wu2017topological,Wu2018,wu2019topological,Wang2020,Li2021quantum,Cai2023,Zeng2023,Park2023,Foutty2024,Guo2025,Xia2025}, where long-period moir\'e patterns induce exotic correlated phenomena.
In addition to these systems constructed by stacking 2D materials, twisted systems of three-dimensional (3D) materials \rev{also possess} various intriguing properties.
For example, twisted 3D graphite~\cite{Koren2016,Li2018,Cea2019,tani2023} (twisted stack of two graphite pieces) exhibits a resonant out-of-plane electronic transport caused by interface-localized states~\cite{tani2023,Cea2019}, and a twisted 3D topological insulator hosts perfect one-dimensional interface states~\cite{Fujimoto2022}.
Also, 3D spiral systems are another example of such systems, which \rev{have} been theoretically and experimentally studied~\cite{Cea2019,Wang2024,wu2020three,Zhao2020,Penghong2022,Chen2025spinless,Ji2024,Phong2025,Lu2024magic}.
 
In Ref.~\cite{tani2023}, we developed a theory for the perpendicular electrical conduction in twisted 3D metals and applied it to twisted 3D graphite as the simplest example.
The present paper aims to extend our theory to treat twisted 3D superconductors.
Here, the central question is what happens if a layered superconductor is twisted with a certain twist angle (see Fig.~\ref{fig:schematic-3dsystem}, a conceptual illustration).
By focusing on a representative layered superconductor NbSe$_2$~\cite{xi2015strongly,xi2016ising,wang2017high,song2024collective,Chen2021twisted,cheung2024coexisting,mchugh2023moire,Doran1978,Lebegue2009,He2018,Habara2021},
previous research~\cite{Yabuki2016,farrar2021superconducting,Jian2022,chen2023interface,Li2023,Ma2025} \rev{has experimentally shown} that a twisted junction between 3D NbSe$_2$ pieces exhibits Josephson effect.
In conventional Josephson junctions (JJs), Josephson effect is caused by a nonsuperconducting barrier (e.g., an insulator) or a weak-link region between two superconductors~\cite{JOSEPHSON1962251,Josephson1974,Likharev1979,Golubov2004}.
{However, in twisted 3D superconductors, there are no such barrier layers, indicating that the junction is purely formed from a twist 
stacking fault. }

{
In this paper, we attempt to clarify how the Josephson effect is induced in twisted interface of superconductors, and how its properties \rev{vary with} changing twist angles.
Toward this goal, we investigate a simple toy model with a Fermi surface around the Brillouin zone (BZ) corner ($K$ point), inspired by NbSe$_2$ that exhibit Fermi surfaces at the BZ corner.
By applying an effective continuum approach \cite{Bistritzer2011,Koshino2018Wannier} to the model, we construct the Bogoliubov–de Gennes (BdG) Hamiltonian of twisted 3D superconductors,
and investigate the properties of the twisted junction.
}

{
As a result of the calculation, we obtain the following findings. The twist causes the two Fermi surfaces around the $K$ point to become misaligned in the in-plane momentum space. This momentum mismatch acts as an effective barrier, leading the system to behave as a JJ. In general, a JJ is characterized by the ratio $J_\mathrm{c} / J_\mathrm{d}$~\cite{sheikhzada2017,plastovets2022,samokhvalov2020,makita2022}, where $J_\mathrm{c}$ is the critical current of the JJ, i.e., the maximum supercurrent that can flow through the junction without a voltage drop. $J_\mathrm{d}$ is the depairing current of the bulk superconducting state, which is independent of the twist interface. It represents the current beyond which Cooper pairs are broken and superconductivity is destroyed in the bulk.
The ratio $J_\mathrm{c} / J_\mathrm{d}$ is generally smaller when a potential barrier in a nonsuperconducting region is high.
In the twisted 3D superconductor, we \rev{demonstrate} that $J_\mathrm{c} / J_\mathrm{d}$ is decreased when the twist angle is increased, indicating that the larger Fermi-surface misalignment naturally works as a higher effective barrier.
Notably, $J_\mathrm{c} / J_\mathrm{d}$ has a finite value (a finite supercurrent flows) even when the Fermi surfaces are completely 
separated in the in-plane momentum.
This is \rev{in sharp contrast} to the normal-state transport,
where the separation of Fermi surfaces eliminates the electric current~\cite{tani2023}.
}

{
Finally, we consider a twisted junction of a 3D superconductor with a Fermi surface centered at the Brillouin zone center ($\Gamma$ point), and demonstrate that the twist has little effect on the electronic transmission across the interface.
We further examine a case with coexisting $\Gamma$- and $K$-pocket Fermi surfaces, as realized in NbSe$_2$, and show that the presence of the $\Gamma$ pocket significantly enhances the ratio $J_\mathrm{c}/J_\mathrm{d}$.
}

{
This paper is organized as follows.
In Sec.~\ref{sec:formulation}, we formulate the BdG Hamiltonian for twisted 3D superconductors using an effective continuum model, and derive the gap equation.
In Sec.~\ref{sec:result}, we present our results on the current-phase relation of twist Josephson junctions and examine its dependence on the twist angle.
In Sec.~\ref{sec:gamma}, we analyze a twisted junction of a 3D superconductor with a Fermi surface centered at the $\Gamma$ point.
Finally, we summarize our findings in Sec.~\ref{sec:conclusion}.
}


\section{Formulation}
\label{sec:formulation}

\subsection{Bulk superconductor model}

\begin{figure}
\begin{center}
   \includegraphics [width=\linewidth]{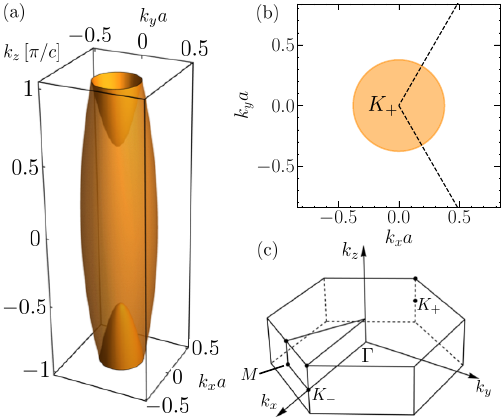}
   \caption{
   (a) Three-dimensional Fermi surface consisting of a cylindrical region with caps at $k_z = \pm \pi/c$.
   The Fermi energy is $E_\mathrm{F} = 0.5 \, \mathrm{eV}$.
   (b) Fermi surface projected onto the $k_x k_y$ plane, appearing as a filled circle. 
   The dashed lines represent the edge of the Brillouin zone.
   (c) Schematic of the 3D Brillouin zone, depicted as a hexagonal prism. 
   High-symmetry points such as $\Gamma$ and $K_\pm$ are indicated.
            }
\label{fig:3d-fs}
\end{center}
\end{figure}

{
We introduce a simple toy model for a layered three-dimensional metal having a Fermi surface near the corner of the Brillouin zone, based on the band structure of graphite~\cite{Slonczewski1958,McClure1957}.
The Fermi surface exists around $K_\xi$ point [see Fig.~\ref{fig:3d-fs}(c)], where the origin of the in-plane wavevector $(k_x, k_y)$ is defined.
The Hamiltonian around $K_\xi$ is given by}
\begin{align}
    \begin{split}
        H^\mathrm{3D}_\xi (\vv{k}_\para, k_z) &= H_\xi(\vv{k}_\para) + T e^{-i k_zc} + T^\dagger e^{i k_zc} - E_\mathrm{F},
        \\
        H_\xi (\vv{k}_\para) &=
        \mqty(0 & -v k_{-} & 0 & 0 \\
        -v k_{+} & 0 & t & 0 \\
        0 & t & 0 & -v k_{-} \\
        0 & 0 & -v k_{+} & 0
        ),
        \\
        T &= 
        \mqty(0 & 0 & 0 & 0 \\
        0 & 0 & t & 0 \\
        0 & 0 & 0 & 0 \\
        0 & 0 & 0 & 0
        ),
    \label{eq:3d-normal-hamil}
    \end{split}
\end{align}
{where $\xi = \pm 1$ is a valley index, $\vv{k}_\para = (k_x, k_y)$ is an in-plane wavevector, $k_\pm = \xi k_x \pm i k_y$, and $k_z$ is an out-of-plane wavenumber. The parameter $c$ is a perpendicular lattice constant,
$t = 0.4 \, \mathrm{eV}$ represents the interlayer hopping, and $v = 0.525 \, \mathrm{eV} \, \mathrm{nm}$ is the Fermi velocity~\cite{Koshino2018Wannier,Moon2013} of a single layer.
The unit structure consists of two layers, and each layer contains two sublattices. The basis of the Hamiltonian is given by the Bloch states $(\ket{A}, \ket{B}, \ket{A'}, \ket{B'})$, where $(A, B)$ and $(A', B')$ correspond to the sublattices of the two different layers in a bilayer unit.
Although this model Hamiltonian is based on AB-stacked graphite, the qualitative properties of the twist Josephson junction are insensitive to the stacking configuration or the inclusion of a mass term, as demonstrated in Appendix~\ref{app:alternative}.
}

Throughout this paper, we employ the Fermi energy $E_\mathrm{F} = 0.5 \, \mathrm{eV}$ for this model.
As shown in Fig.~\ref{fig:3d-fs}(a), the shape of the Fermi surface is an open cylinder with caps around $k_z = \pm \pi/c$.
Here, the parameter $a$ is \rev{an} in-plane lattice constant.
When we project the Fermi surface onto the $k_x k_y$ plane, its projection simply forms a filled circle [Fig.~\ref{fig:3d-fs}(b)].
Figure~\ref{fig:3d-fs}(c) illustrates the 3D Brillouin zone, which has the shape of a hexagonal prism.

When we consider the superconducting order in this system, the Bogoliubov–de Gennes (BdG) Hamiltonian in Nambu space is written as
\begin{equation}
\begin{split}
\label{eq:3d-bdg}
    H^\mathrm{3D \text{-} BdG}_\xi (\vv{k}_\para, k_z) &= \mqty(
    H^\mathrm{3D}_\xi (\vv{k}_\para, k_z) & \underline{\Delta} \\
    \underline{\Delta}^* & -H^{\mathrm{3D}*}_{-\xi} (-\vv{k}_\para, -k_z)
    ), 
    \\
    \underline{\Delta} &= \mqty(
    \Delta_A & & & \\
    & \Delta_B & & \\
    & & \Delta_{A'} & \\
    & & & \Delta_{B'}
    ),
\end{split}
\end{equation}
where $\Delta_X$ is \rev{the} $s$-wave order parameter for $X = A, B, A', B'$.
Expressing this complex number as $\Delta_X = |\Delta_X| e^{i \varphi_X}$, we identify $|\Delta_X|$ as the amplitude and $\varphi_X$ as the superconducting phase.
Note that the superconducting order parameter $\underline{\Delta}$ hybridizes the opposite valleys $\xi$ and $-\xi$.



\subsection{Twisted 3D superconductor}

\begin{figure}
\begin{center}
   \includegraphics [width=0.8\linewidth]{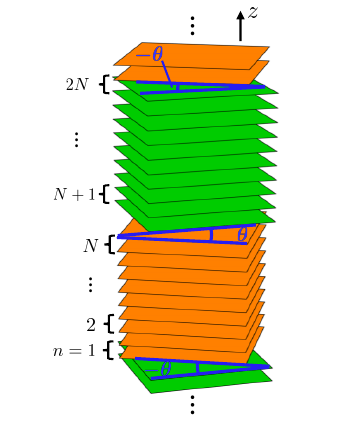}
   \caption{
   Side view of a twisted three-dimensional superconductor, where the lower (upper) slabs are shown by the orange (green) layers.
   The periodic boundary condition is employed for the calculation of the perpendicular supercurrent.
   The system contains $2N$ bilayer units, which are labeled by an integer $n$ (see the text).
   The twisted interfaces are located between $n = N$ and $N+1$ with the angle $\theta$, and between $n = 1$ and $n = 2N$ with the angle $-\theta$.
   }
\label{fig:sideview}
\end{center}
\end{figure}

{
We consider a twisted interface as illustrated in Fig.~\ref{fig:sideview}, where slabs of the three-dimensional superconductor introduced in the previous section are stacked with a relative twist angle $\theta$.
Specifically, we consider lower ($l=1$) and upper ($l=2$) slabs, each of which includes $N$ bilayer units ($2N$ layers).
The in-plane lattice orientations of the two slabs are relatively rotated by $\theta$.
We label bilayer units by $n = 1, 2, \cdots, N$ for the lower slab, and $n = N+1, N+2, \cdots, 2N$ for the upper slab.
For the calculation of the perpendicular supercurrent, we employ the periodic boundary condition between the bilayer units $n=1$ and $n=2N$.
The twisted interfaces exist between $n=N$ and $n=N+1$ (twist angle $\theta$), and also between $n=1$ and $n=2N$ (twist angle $-\theta$).
We label sublattices in the $n$th bilayer unit as $A_n, B_n, A'_n, B'_n$.
}

\begin{figure}[b]
\begin{center}
   \includegraphics [width=\linewidth]{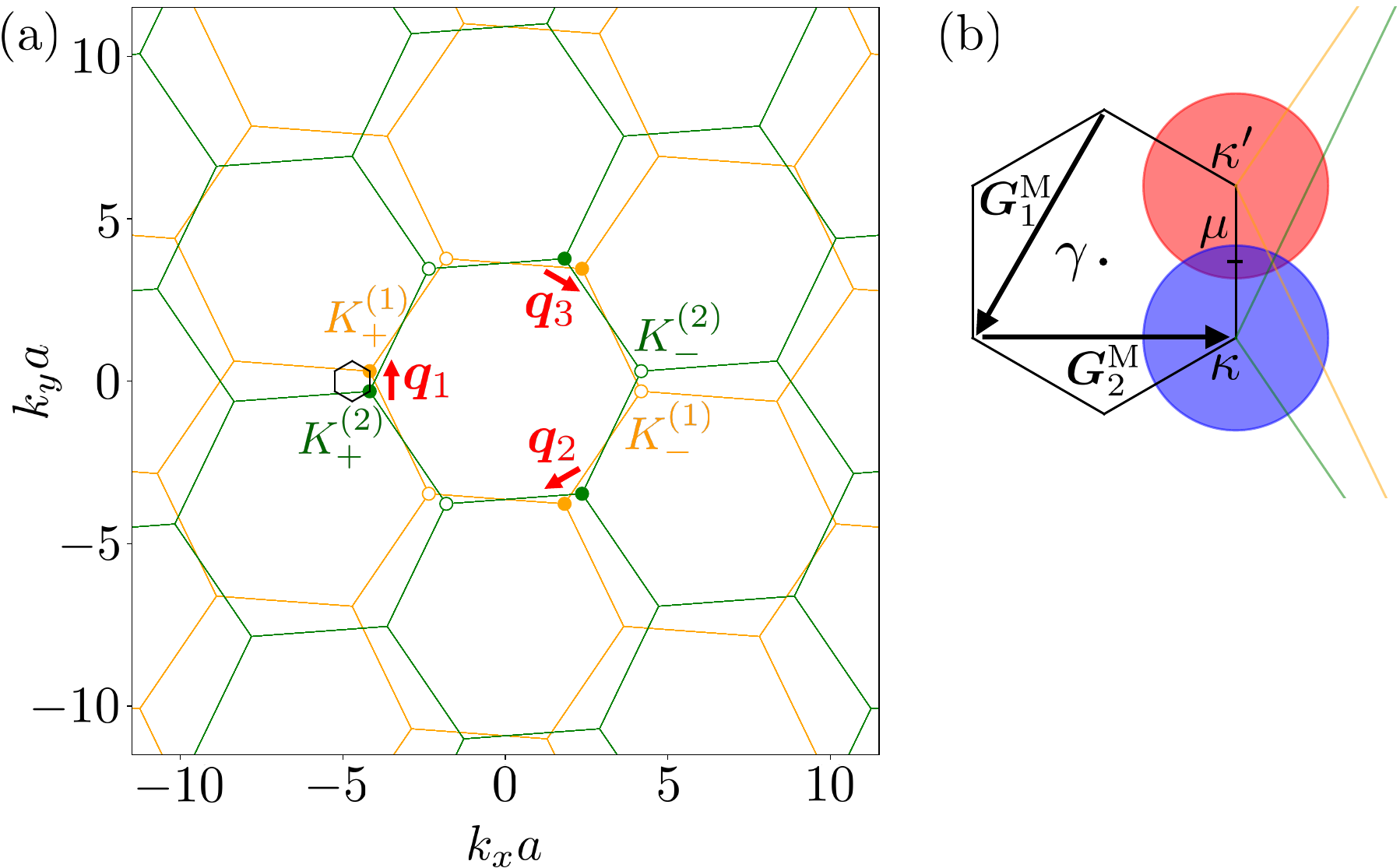}
   \caption{
   (a)~Brillouin zones of the lower (yellow) and upper (green) honeycomb lattices in the extended-zone scheme. 
   The coupling wavevectors $\vv{q}_1$, $\vv{q}_2$, and $\vv{q}_3$ are indicated by red arrows (see the text). 
   A moir\'{e} Brillouin zone is defined by a small black hexagon near $\vv{q}_1$, which is magnified in (b).
   (b)~Misaligned Fermi surfaces of the lower and upper 3D metals around the $K_+$ point are illustrated as red and blue circles, respectively. 
   With the moir\'e Brillouin zone (black hexagon), the moir\'{e} reciprocal vectors $\vv{G}^\mathrm{M}_1$, $\vv{G}^\mathrm{M}_2$ and high-symmetry points are shown as well.
            }
\label{fig:moirebz}
\end{center}
\end{figure}

{
The Brillouin zone (BZ) corners of the slab $l$ \rev{are} given by $\vv{K}_\xi^{(l)} = R^{(l)} \vv{K}_\xi$,  where
\begin{equation}
    R^{(1)} = R(-\theta/2),\quad
    R^{(2)} = R(+\theta/2),
\end{equation}
and $R(\alpha)$ is the rotation matrix by an angle of $\alpha$.
We define the interlayer displacement vectors of the three BZ corners as
$\vv{q}_1 = \vv{K}_+^{(1)} - \vv{K}_+^{(2)}, \bm{q}_2 = R(120^\circ)\bm{q}_1,  \, \bm{q}_3= R(-120^\circ)\bm{q}_1$
as shown in Fig.~\ref{fig:moirebz}(a).
The original Brillouin zone [Fig.~\ref{fig:3d-fs}(c)] is folded into the moir\'{e} Brillouin zone (MBZ), which is illustrated as a small black hexagon in Fig.~\ref{fig:moirebz}(a).
As illustrated in Fig.~\ref{fig:moirebz}(b), the moir\'e reciprocal lattice vectors to span the MBZ are given by $\vv{G}^\mathrm{M}_1 = \vv{q}_2 - \vv{q}_1$ and $\vv{G}^\mathrm{M}_2 = \vv{q}_3 - \vv{q}_2$.
Because of the rotation in wavevector space, the Fermi surfaces are also rotated and become misaligned relative to each other.
Figure \ref{fig:moirebz}(b) illustrates the rotated Fermi surfaces of the lower (upper) parts as the red (blue) filled circle, respectively.
}

{
We employ an effective continuum model of the entire system analogous to that for twisted bilayer graphene~\cite{Bistritzer2011,Koshino2018Wannier}.
In a basis ordered as 
$(\cdots;\ket*{A_n},\ket*{B_n},\ket*{A'_n},\ket*{B'_n};\ket*{A_{n+1}},\ket*{B_{n+1}},$ 
$\ket*{A'_{n+1}},\ket*{B'_{n+1}};\cdots)$, 
the normal-state Hamiltonian is given by
}


\begin{widetext}
\begin{equation}
\begin{split}
\mathcal{H}_\xi (\vv{k}_\para) =
    \left(
    \begin{array}{ccccc|ccccc}
    H^{(1)} & T^{(1)\dagger} & & & & & & & & T'_{\mathrm{int}}
    \\
    T^{(1)} & H^{(1)} & T^{(1)\dagger} & & & & &
    \\
    & \ddots & \ddots & \ddots & & & & & 
    \\
    & & T^{(1)} & H^{(1)} & T^{(1)\dagger} & & & & 
    \\
    & & & T^{(1)} & H^{(1)} & T_{\mathrm{int}}^{\dagger} & & &
    \\[2pt]
    \hline
    & & & & T_{\mathrm{int}} & H^{(2)} & T^{(2)\dagger} & &
    \\
    & & & & & T^{(2)} & H^{(2)} & T^{(2)\dagger} & 
    \\
    & & & & & & \ddots & \ddots & \ddots 
    \\
    & & & & & & & T^{(2)} & H^{(2)} & T^{(2)\dagger}
    \\
    T'^{\dagger}_{\mathrm{int}} & & & & & & & & T^{(2)} & H^{(2)}
    \\
    \end{array}\right) - E_\mathrm{F},
\label{eq:twist-normal-hamil}
\end{split}
\end{equation}
\end{widetext}
where $4\times4$ matrices $H^{(l)}$ and $T^{(l)}$ are defined as
\begin{equation}
H^{(l)} = H_\xi[(R^{(l)})^{-1} \bm{k}_\para],
\quad
T^{(l)} = T.
\end{equation}
The matrices $H_\xi$ and $T$ are defined by Eq.~\eqref{eq:3d-normal-hamil}.

The symbol $T_{\mathrm{int}}$ denotes the interlayer Hamiltonian matrix for the central twisted interface ($n=N$ and $n=N+1$ bilayer units), which is given by \cite{Bistritzer2011,Koshino2018Wannier}
\begin{equation}
    \begin{split}
        & T_{\mathrm{int}} = \mqty(0 & T_{\mathrm{int}}^{2\times 2} \\ 0 & 0), \quad
        T_{\mathrm{int}}^{2\times 2} (\vv{r}_\para) = \sum_{j=1}^{3} U_j e^{i \xi \vv{q}_j \vdot \vv{r}_\para},
        \\
        & U_1 = \mqty(u&u'\\u'&u), \,
        U_2 = \mqty(u&u'\omega^{-\xi}\\u'\omega^{\xi}&u), \,
        U_3 = \mqty(u&u'\omega^{\xi}\\u'\omega^{-\xi}&u),
    \end{split}
\label{eq:interlayer}
\end{equation}
where $\omega=e^{2\pi i/3}$, $u=0.0797~\mathrm{eV}$, and $u'=0.0975~\mathrm{eV}$~\cite{Koshino2018Wannier}.
Since we employ the periodic boundary condition, the coupling matrix between the boundary bilayer units ($n=1$ and $n=2N$) is written as
\begin{equation}
    T_{\mathrm{int}}' = \mqty(0 & T_{\mathrm{int}}'^{2\times 2} \\ 0 & 0), \quad
        T_{\mathrm{int}}'^{2\times 2} (\vv{r}_\para) = \sum_{j=1}^{3} U_j e^{i \xi (-\vv{q}_j) \vdot \vv{r}_\para}.
\label{eq:interlayer-pbc}
\end{equation}
Starting from a lower-part Bloch state of the wavevector $\vb*{k}_{\para}$, the interlayer Hamiltonian $T_{\mathrm{int}}$ and $T_{\mathrm{int}}'$ hybridize a set of wavenumbers of the same valley,
\begin{equation}
\begin{split}
\label{eq:q}
  &  \bm{k}^{(1)}_{\para}(m_1,m_2) = 
  \xi\bm{q}_1 + \bm{k}_{\para} + m_{1}\bm{G}^{\mathrm{M}}_{1} + m_{2}\vb*{G}^{\mathrm{M}}_{2},
    \\
  &   \bm{k}^{(2)}_{\para}(m_1,m_2) =  
  \bm{k}_{\para} + m_{1}\bm{G}^{\mathrm{M}}_{1} + m_{2}\vb*{G}^{\mathrm{M}}_{2},
\end{split}
\end{equation} 
of the upper and lower parts, respectively ($m_1, m_2$: integers).
To express the Hamiltonian as a finite-sized matrix, we consider a finite set of wavenumbers
inside a certain cutoff circle $|\bm{k}^{(l)}_{\para}| \leq k_{\mathrm{c}}$.
Note that the in-plane wavevector $\vb*{k}_{\para}$ is a parameter that varies within the MBZ (Fig.~\ref{fig:moirebz}).
In this representation, $H^{(l)}$, $T^{(l)}$, $T_{\mathrm{int}}$, and $T'_{\mathrm{int}}$ in 
Eq.~\eqref{eq:twist-normal-hamil}
are $4N^{(l)}_q\times 4N^{(l)}_q$ matrices, where
$N^{(l)}_q$ denotes the number of distinct wavenumbers in the set of $\{\bm{k}^{(l)}_{\para} \}$, and 
the factor 4 accounts for the sublattices ($A,B,A',B'$).
The matrices $H^{(l)}$ and $T^{(l)}$ are diagonal with respect to the label $(m_1,m_2)$, whereas only the \rev{matrices} $T_{\mathrm{int}}$ and $T'_{\mathrm{int}}$ \rev{hybridize} different $(m_1,m_2)$ components.

Similar to the 3D case [Eq.~\eqref{eq:3d-bdg}], the BdG Hamiltonian in Nambu space is constructed from the normal-state Hamiltonian [Eq.~\eqref{eq:twist-normal-hamil}].
Here we assume an $s$-wave superconducting order.
Noting that the Cooper pair consists of electrons having $(\xi, \vv{k}_\para)$ and $(-\xi, -\vv{k}_\para)$ (opposite valleys and wavevectors), the BdG Hamiltonian is written as
\begin{widetext}
\begin{gather}
\label{eq:twist-bdg-hamil}
    \mathcal{H}_\xi^\mathrm{BdG}(\vv{k}_\para) =
    \mqty(
    \mathcal{H}_\xi(\vv{k}_\para) & \underline{\Delta}(\vv{r}_\para) \\
    \underline{\Delta}^{*} (\vv{r}_\para) & -\mathcal{H}_{-\xi}^{*} (-\vv{k}_\para)
    ),
    \\
    \underline{\Delta}(\vv{r}_\para) =
    \left(
    \begin{array}{ccccc|ccccc}
    \underline{\Delta}_{1}^{4\times4} & & & & & & & & & 
    \\
    & \underline{\Delta}_{2}^{4\times4} & & & & & &
    \\
    & & \ddots & & & & & & 
    \\
    & & & \underline{\Delta}_{N-1}^{4\times4} & & & & & 
    \\
    & & & & \underline{\Delta}_{N}^{4\times4} & & & &
    \\[2pt]
    \hline
    & & & & & \underline{\Delta}_{N+1}^{4\times4} & & &
    \\
    & & & & & & \underline{\Delta}_{N+2}^{4\times4} & & 
    \\
    & & & & & & & \ddots &
    \\
    & & & & & & & & \underline{\Delta}_{2N-1}^{4\times4} &
    \\
    & & & & & & & & & \underline{\Delta}_{2N}^{4\times4}
    \\
    \end{array}\right),
    \\
    \underline{\Delta}_n^{4\times4} (\vv{r}_\para) = 
    \mqty(
    \Delta_{A_n} & & &
    \\
    & \Delta_{B_n} & &
    \\
    & & \Delta_{A'_n} &
    \\
    & & & \Delta_{B'_n}
    ),
\end{gather}
\end{widetext}
{where the order parameter matrix $\underline{\Delta} (\vv{r}_\para)$ is diagonal with respect to both the bilayer unit index ($n = 1, 2, \cdots, 2N$) and the sublattice index ($X=A, B, A', B'$); each component is denoted by $\Delta_{X_n} (\vv{r}_\para)$.
However, $\Delta_{X_n} (\vv{r}_\para)$ connects different moir\'e wavevectors~[Eq.~\eqref{eq:q}] since a moir\'e-scale modulation is allowed in its \rev{spatial} distribution.
In addition, we assume \rev{an} $s$-wave order parameter, and therefore $\Delta_{X_n} (\vv{r}_\para)$ is independent of the in-plane wavevector $\vv{k}_\para$.}

The BdG equation to be solved is
\begin{equation}
\label{eq:twist-bdg-eq}
    \mathcal{H}_\xi^\mathrm{BdG}(\vv{k}_\para) \,
    \mqty(\vv{u}_{\xi \vv{k}_\para}^{\nu} \\ \vv{v}_{\xi \vv{k}_\para}^{\nu})
    = E^{\nu}_{\xi \vv{k}_\para} \, \mqty(\vv{u}_{\xi \vv{k}_\para}^{\nu} \\ \vv{v}_{\xi \vv{k}_\para}^{\nu}),
\end{equation}
where $\nu$ is an index designating the eigenstates, $E^{\nu}_{\xi \vv{k}_\para}$ is an energy eigenvalue, and $\vv{u}_{\xi \vv{k}_\para}^{\nu} (\vv{r}_\para)$ [$\vv{v}_{\xi \vv{k}_\para}^{\nu} (\vv{r}_\para)$] is the $8N$-component wavefunction corresponding to the electron (hole) part of Nambu space.
Explicitly, they are expressed as $\vv{u} (\vv{r}_\para) = ( u_{A_1}(\vv{r}_\para), u_{B_1}(\vv{r}_\para), \cdots , u_{B'_{2N}}(\vv{r}_\para))$; $\vv{v}(\vv{r}_\para)$ is defined in the same manner.
A detailed derivation of the BdG equation is provided in Appendix \ref{app:valley}.

Using the wavefunction, the gap equation determining the $s$-wave order parameter is written as
\begin{equation}
\label{eq:gapeq-realspace}
    \Delta_{X_n} (\vv{r}_\para) =
    -V \sum_{\alpha}
    v_{X_n}^{\alpha*} (\vv{r}_\para)
    u_{X_n}^{\alpha} (\vv{r}_\para)
    f(E^\alpha),
\end{equation}
{where $X_n = A_n, B_n, A'_n, B'_n$ is the sublattice index,
$\alpha$ is a joint index for $(\nu,\xi,\vv{k}_\para)$, 
the $V>0$ is the strength of an electron-electron interaction,
and $f(E)$ is the Fermi distribution function.
The summation  in $\alpha$ is defined as
\begin{equation}
    \sum_\alpha = \sum_{\xi=\pm} \sum_{\nu}
    \int
    \frac{\dd^2 \vv{k}_\para}{(2\pi)^2}.
\end{equation}
}
In the present paper, all calculations are performed at zero temperature.
Here, the valley index $\xi = \pm$ is summed over since it serves merely as a label specifying the states (a more detailed discussion is presented in Appendix~\ref{app:valley}).
The self-consistent solution of the system is obtained by recursively solving Eqs.~\eqref{eq:twist-bdg-eq} and \eqref{eq:gapeq-realspace}. 
To ensure convergence, we introduce an energy cutoff $\delta E$ which is larger than the superconducting gap.
The integration is limited to wavevectors $\vv{k}_\para$ satisfying $|E^\nu_{\xi\vv{k}_\para}| < \delta E$.
Therefore, we first fix the parameters $(V, \delta E)$ and then obtain the self-consistent order parameter $\underline{\Delta} (\vv{r}_\para)$.

The order parameter $\Delta_{X_n} (\vv{r}_\para)$ has off-diagonal parts that connect different wavevectors $(m_1, m_2)$~[see Eq.~\eqref{eq:q}].
In general, it can be written as
\begin{equation}
\label{eq:fourier-delta}
    \Delta_{X_n} (\vv{r}_\para) = \sum_{\vv{q}} 
    \Delta_{X_n \vv{q}} e^{i \vv{q} \vdot \vv{r}_\para},
\end{equation}
where the summation is taken over indices $(m_1, m_2)$ corresponding to $\vv{q} = m_{1}\bm{G}^{\mathrm{M}}_{1} + m_{2}\vb*{G}^{\mathrm{M}}_{2}$.
Each Fourier component $\Delta_{X_n \vv{q}}$ represents a moir\'e-scale modulation in the in-plane direction.
Similarly, the wavefunction is expanded as
\begin{equation}
\label{eq:fourier-wavefunc}
    \begin{split}
    &    u_{X_n}^{\alpha} (\vv{r}_\para) = \sum_{\vv{q}} 
    u_{X_n \vv{q}}^{\alpha} \,
    e^{i \vv{q} \vdot \vv{r}_\para},
    \\
    &   v_{X_n}^{\alpha} (\vv{r}_\para) = \sum_{\vv{q}} 
    v_{X_n \vv{q}}^{\alpha} \,
    e^{i \vv{q} \vdot \vv{r}_\para}.
    \end{split}
\end{equation}
From Eqs.~\eqref{eq:gapeq-realspace}, \eqref{eq:fourier-delta}, and \eqref{eq:fourier-wavefunc}, the Fourier component of the order parameter is given by
\begin{equation}
   \Delta_{X_n \vv{q}} =
    -V \sum_{\alpha} \sum_{\vv{q}'}
    v_{X_n \vv{q}'}^{\alpha *}
    u_{X_n, \vv{q} + \vv{q}'}^{\alpha}
    f(E^{\alpha}).
\end{equation}

To examine how the superconducting phase varies along the perpendicular direction, we enforce a certain phase difference between the bottommost bilayer unit ($n=1$) and topmost bilayer unit ($n=2N$) by introducing an appropriate boundary condition as follows.
In the BdG Hamiltonian [Eq.~\eqref{eq:twist-bdg-hamil}], $\mathcal{H}_\xi (\vv{k}_\para)$ represents the electron component in Nambu space.
In this matrix, we replace the boundary coupling matrix $T_\mathrm{int}'^{\dagger}$ with
\begin{equation}
    T_\mathrm{int}'^{\dagger} \rightarrow T_\mathrm{int}'^{\dagger} e^{i \delta\varphi / 2},
\end{equation}
where $\delta\varphi$ is the desired phase difference.
This replacement corresponds to the condition
\begin{equation}
    \mqty(u_{X_{2N+1}}^\alpha (\vv{r}_\para) \\ v_{X_{2N+1}}^\alpha (\vv{r}_\para)) =
    \mqty(e^{i\delta\varphi/2} & 0 \\ 0 & e^{-i\delta\varphi/2})
    \,
    \mqty(u_{X_1}^\alpha (\vv{r}_\para) \\ v_{X_1}^\alpha (\vv{r}_\para)),
\end{equation}
leading to
\begin{equation}
    \Delta_{X_{2N+1}} = \Delta_{X_{1}} e^{i\delta\varphi}.
\end{equation}
Thus, the phase difference is fixed by the relation
\begin{equation}
    \varphi_{X_{2N+1}} - \varphi_{X_{1}} = \delta\varphi.
\end{equation}

{
The averaged current density along \rev{the} $z$ direction is calculated by
\begin{gather}
\label{eq_J}
J = \frac{1}{S_\mathrm{M}}
    \int_{S_\mathrm{M}} \dd^2 \vv{r}_\para \,
    J_n(\vv{r}_\para),
    \\
\begin{split}
\label{eq:current-realspace}
    & J_n(\vv{r}_\para) = \frac{-2e}{\hbar} \times
    \\ 
    & \Im \sum_{\alpha} \sum_{X, \tilde{X}}
    \biggl[
    u_{X_{n+1}}^{\alpha*}(\vv{r}_\para)
    [\mathcal{H}_\xi(\vv{k}_\para)]_{X_{n+1}, \tilde{X}_n}
    u_{\tilde{X}_{n}}^{\alpha}(\vv{r}_\para) 
    f(E^\alpha)
    \\
    & \hspace{1cm} + v_{X_{n+1}}^{\alpha}(\vv{r}_\para)
    [\mathcal{H}_\xi(\vv{k}_\para)]_{X_{n+1}, \tilde{X}_n}
    v_{\tilde{X}_{n}}^{\alpha*}(\vv{r}_\para)
    f(-E^\alpha)
    \biggr],
\end{split}
\end{gather}
where $[\mathcal{H}_\xi(\vv{k}_\para)]_{X_{n+1}, \tilde{X}_n}$ is the $4\times4$ partial block connecting $X_{n+1} = (A_{n+1}, B_{n+1}, A'_{n+1}, B'_{n+1})$ and $\tilde{X}_n = (A_n, B_n, A'_n, B'_n)$.
\rev{Here, $\alpha$ is the joint index for $(\nu,\xi,\vv{k}_\para)$, and $(\xi,\vv{k}_\para)$ in the Hamiltonian $\mathcal{H}_\xi(\vv{k}_\para)$ is a component of $\alpha$.}
\rev{In Eq.~\eqref{eq_J}, the current $J_n(\vv{r}_\para)$ is averaged over the moir\'e unit cell $S_\mathrm{M}$.}
Note that $J$ is independent of $n$ because of current conservation in the perpendicular direction.
A derivation of Eq.~\eqref{eq:current-realspace} is given in Appendix \ref{app:supercurrent}.
We also define the averaged order parameter as 
\begin{gather}
    \overline{\Delta}_{X_n} = \frac{1}{S_\mathrm{M}}
    \int_{S_\mathrm{M}} \dd^2 \vv{r}_\para \,
    \Delta_{X_n} (\vv{r}_\para).
\end{gather}
}


\section{Twist Josephson junction}
\label{sec:result}



Based on the formulation established in Sec.~\ref{sec:formulation}, we examine the properties of the twisted 3D superconductor.
Throughout the paper, we consider the system with the periodic boundary condition with $N = 10$ ($40$ layers in total).
 {First, we investigate
the spatial variation of the superconducting order parameter along the out-of-plane direction.
In our numerical calculation, a phase twist of the boundary condition $\delta \varphi$ is a given parameter, 
where a change in $\delta\varphi$ corresponds to varying the supercurrent $J$ in the perpendicular direction.
For a given $\delta\varphi$, we self-consistently solve the BdG equation [Eq.~\eqref{eq:twist-bdg-eq}] and the gap equation [Eq.~\eqref{eq:gapeq-realspace}].
\rev{This corresponds to a situation where a finite supercurrent is applied to the system, rather than the global equilibrium state. 
The obtained current-carrying state should thus be regarded as a possible steady state sustained by an external phase bias.
}

The calculation is initiated with some specific spatial profile for $\Delta_{X_n}$, and iterated until a converged solution is reached.
We have confirmed that the final self-consistent state is independent of the initial choice for $\Delta_{X_n}$.
Here we employ $V = 1 \, \mathrm{eV}$ and $\delta E = 0.1 \, \mathrm{eV}$ to obtain the bulk superconducting gap of 
$\Delta_{\rm bulk} \equiv \Delta_{A} = \Delta_{B'} \approx 20 \, \mathrm{meV}$.
Here, $A$ and $B'$ sublattices predominantly contribute to the Fermi surface, and therefore to the superconducting states.
}

\begin{figure}
\begin{center}
   \includegraphics [width=0.8\linewidth]{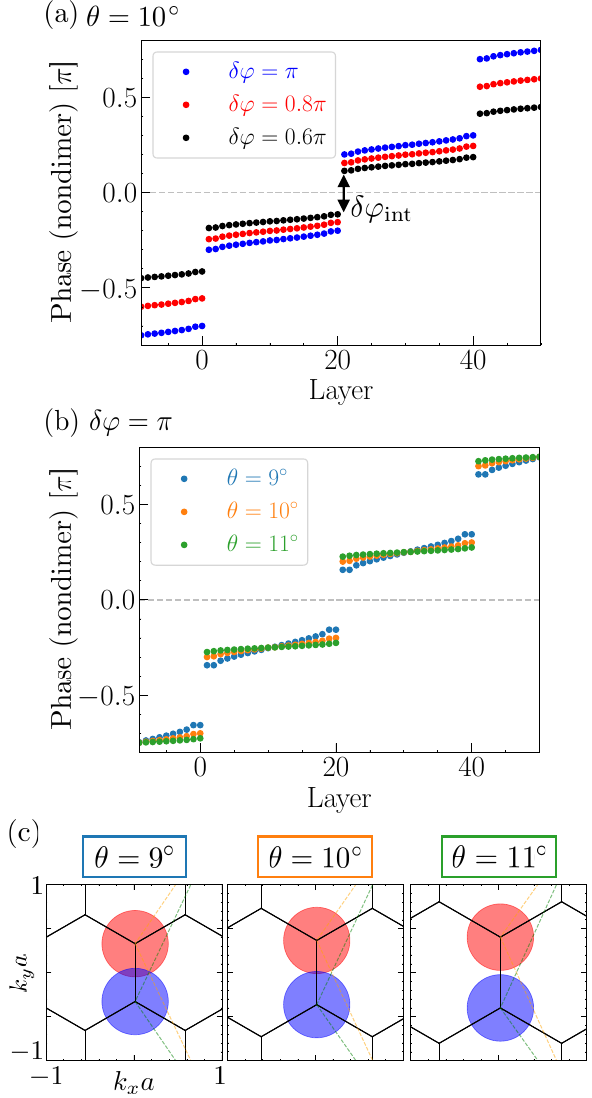}
   \caption{
   (a)~Variation of the superconducting phase (on the nondimer sites) along the stacking direction.
   The phase twist $\delta\varphi$ is varied, while the twist angle is fixed to $\theta=10^\circ$.
   $\delta\varphi_\mathrm{int}$ denotes the phase jump at the twisted interface (between layers $20n$ and $20n+1$).
   (b)~Similar plots for various twist angles $\theta$, \rev{with $\delta\varphi$ fixed at $\pi$.}
   (c)~Misaligned Fermi surfaces projected onto the $k_x k_y$ plane for the three twist angles.  
   The red (blue) circle represents the Fermi surface of the lower (upper) part of the system.
            }
            \label{fig:phase_jump}
 \end{center} 
 \end{figure}

{
In Fig.~\ref{fig:phase_jump}(a), we illustrate the 
variation of the superconducting phase 
$\overline{\varphi}_{X_n} \equiv \arg \overline{\Delta}_{X_n}$ along the stacking direction ($z$ direction), in a twisted 3D superconductor with twist angle $\theta = 10^\circ$
with different values of $\delta\varphi$.
The horizontal axis represents the layer index, where 
$\overline{\varphi}_{A_1}, \overline{\varphi}_{B'_1}, \cdots, \overline{\varphi}_{A_{2N}}, \overline{\varphi}_{B'_{2N}}$
(the phases on the nondimer sites) are plotted for layers $1, 2, \cdots, 4N\ (=40)$.
Here, $\delta\varphi$ corresponds to the total phase difference across a single superlattice period (from layer 1 to 40),
and the plot is extended beyond the period using the imposed boundary condition.
The superconducting phase exhibits a jump $\delta\varphi_\mathrm{int}$ at the twisted interface (between layers $20n$ and $20n+1$),
while it varies smoothly in the bulk regions.
As $\delta\varphi$ increases, both the phase jump at the interface and the phase gradient in the bulk region increase proportionally.
}

{
Figure~\ref{fig:phase_jump}(b) shows similar plots for different twist angles, $\theta = 9^\circ$, $10^\circ$, and $11^\circ$, with $\delta\varphi$ fixed at $\pi$. As the twist angle increases, the phase change becomes more localized at the interface, approaching the weak-link limit, where the entire phase jump is concentrated at the interface. In contrast, as the twist angle decreases, the system approaches the strong-link limit, where the phase gradient is nearly uniform throughout the entire system, as in the bulk.  
This variation with twist angle corresponds directly to the overlap of the momentum range [see Fig.~\ref{fig:phase_jump}(c)]: as $\theta$ increases, the Fermi surfaces of the lower part (layers 1 to 20) and the upper part (layers 21 to 40) \rev{are becoming} increasingly misaligned, \rev{and} completely disconnected at $\theta \sim 10^\circ$. Accordingly, electronic transfer across the interface is significantly reduced, leading to a crossover from the strong-link to the weak-link regime.
}

\begin{figure}
\begin{center}
   \includegraphics [width=0.75\linewidth]{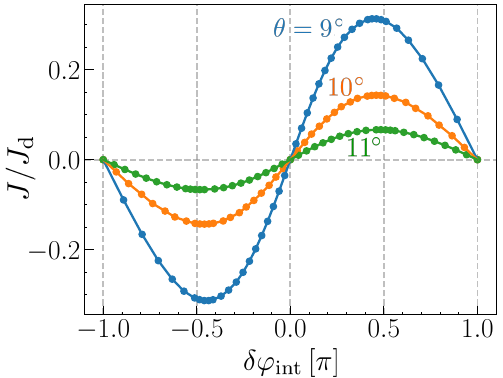}
   \caption{
   Current-phase relations $J = J(\delta\varphi_\mathrm{int})$ for various twist angles $\theta$.  
   The supercurrent is normalized by the depairing current $J_\mathrm{d}$ (see the main text).
            }
            \label{fig:CPR}
 \end{center}
 \end{figure}

{
We can show that the perpendicular electric current $J$ [Eq.~\eqref{eq_J}] calculated for the given system is solely dependent 
on the interface phase jump $\delta\varphi_{\rm int}$,
regardless of the size $N$ chosen for the numerical calculation.
Figure~\ref{fig:CPR} plots $J$ [Eq.~\eqref{eq_J}]
as a function of $\delta\varphi_{\rm int}$, for various twist angles.
In Fig.~\ref{fig:CPR}, $J$ is significantly reduced with increasing twist angle, reflecting the reduction of Fermi surface overlap shown in Fig.~\ref{fig:phase_jump}(c).
The critical Josephson current $J_\mathrm{c}$ is defined as the maximum value of the current $J$:
\begin{equation}
    J_\mathrm{c} = \max_{\delta\varphi_\mathrm{int}} J (\delta\varphi_\mathrm{int}).
\end{equation}
}

{
In Fig.~\ref{fig:CPR}, the supercurrent $J$ is normalized by the depairing current $J_\mathrm{d}$, which is the Josephson critical current of the 3D bulk system without a twisted interface, i.e., \rev{the current at which} Cooper pairs are depaired
when the applied current exceeds $J_\mathrm{d}$.
Formally, $J_\mathrm{d}$ is defined in a similar manner to $J_\mathrm{c}$,
\begin{equation}
    J_\mathrm{d} = \max_{\delta\varphi_\mathrm{bulk}} J_\mathrm{bulk}(\delta\varphi_\mathrm{bulk}),
\label{eq:jd-definition}
\end{equation}
where $J_\mathrm{bulk}$ is the Josephson current in the nontwisted bulk 3D system.
$\delta\varphi_\mathrm{bulk}$ is the phase twist between adjacent bilayer units.
A detailed procedure to calculate $J_\mathrm{bulk}$ is provided in Appendix~\ref{app:depairing}.
}
{
From the value of $J_\mathrm{c} / J_\mathrm{d}$, one can infer the effective coupling strength between the lower and upper parts of the system, as well as the nature of the Josephson junction.  
When $J_\mathrm{c} / J_\mathrm{d} \ll 1$, only a very weak supercurrent can flow compared to the corresponding bulk system, indicating that the system is close to the weak-link regime.  
On the other hand, when $J_\mathrm{c} / J_\mathrm{d}$ approaches unity, the junction supports a robust supercurrent comparable to the bulk depairing current, signifying a high-$J_\mathrm{c}$ Josephson junction.
}

 
\begin{figure}
\begin{center}
   \includegraphics [width=\linewidth]{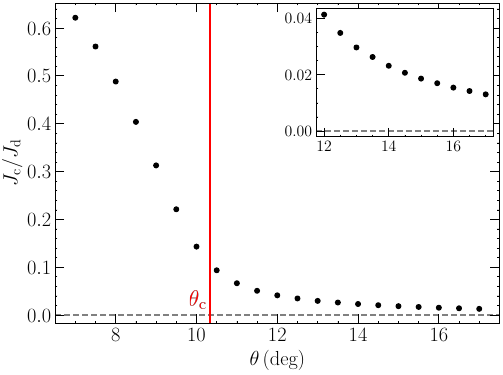}
   \caption{
   Twist-angle dependence of the Josephson critical current normalized by the depairing current $J_\mathrm{c} / J_\mathrm{d}$.  
   The vertical red line indicates the critical angle $\theta_\mathrm{c}$, beyond which the Fermi surfaces are disconnected.
            }
            \label{fig:jcjd-theta}
 \end{center}
 \end{figure}
{
Figure \ref{fig:jcjd-theta} plots $J_\mathrm{c} / J_\mathrm{d}$ as a function of the twist angle $\theta$.
The vertical red line represents the critical angle \rev{$\theta_\mathrm{c}\approx 10.3^\circ$}, where the top and bottom Fermi surfaces become completely separated.
Overall, as the twist angle increases from $\theta=7^\circ$, $J_\mathrm{c} / J_\mathrm{d}$ drops in accordance with the decrease in the Fermi-surface overlap.
Notably, a nonzero value of $J_\mathrm{c} / J_\mathrm{d}$ persists even for $\theta > \theta_\mathrm{c}$,  
i.e., a finite supercurrent can flow despite the Fermi surfaces being completely separated in momentum space,  
and it gradually decays as $\theta$ increases.
}

{
This behavior stands in sharp contrast to a twisted 3D interface of normal metals, where the perpendicular conduction simply vanishes when the Fermi surface overlap is absent~\cite{tani2023}.  
Specifically, in a twisted junction of 3D metals, the electrical conductivity is given by the sum of momentum-resolved conductance $G(\vv{k}_\parallel)$ at each in-plane wavevector $\vv{k}_\parallel$~\cite{tani2023}.  
Because of the in-plane moiré periodicity, $\vv{k}_\parallel$ must be conserved during scattering at the twisted interface.  
As a result, when $\theta > \theta_\mathrm{c}$, $G(\vv{k}_\parallel) = 0$ for all $\vv{k}_\parallel$, leading to a complete suppression of perpendicular conduction.
In contrast, twisted 3D superconductors exhibit a qualitatively different behavior, since the current carriers are not single electrons but Cooper pairs.  
Typically, a Cooper pair consists of two electrons with in-plane momenta $(\vv{k}_\parallel, -\vv{k}_\parallel)$.  
After scattering at the interface, the pair can in principle transition to a different state $(\vv{k}'_\parallel, -\vv{k}'_\parallel)$ with arbitrary $\vv{k}'_\parallel$, as long as the total momentum remains zero.  
This momentum flexibility allows supercurrent to flow across the interface even when the Fermi surfaces of the two superconductors are completely separated.
}
\rev{It is important to clarify that this anomalous Cooper pair transport is a direct consequence of the standard contact electron-electron interaction model (see Appendix~\ref{app:valley}).
In this framework, the zero interaction range in real space ($\Delta r = 0$) translates to a constant, $k$-independent interaction in momentum space. 
As a result, the gap function $\Delta$ in the BdG Hamiltonian receives uniform contributions from all quasiparticle states across the entire Brillouin zone [Eq.~\eqref{eq:gapeq-realspace}].
This nonlocal coupling in $k$ space facilitates the scattering of Cooper pairs between states $(\bm{k}_\parallel, -\bm{k}_\parallel)$ and $(\bm{k}_\parallel', -\bm{k}_\parallel')$ even when their respective Fermi surfaces are spatially separated in the Brillouin zone.
In contrast, if a finite interaction range ($\Delta r > 0$) were introduced, the $k$-space interaction would become localized over a range $\sim 1/\Delta r$.
The order parameter $\Delta (\bm{k}_\parallel)$, which now depends on $\vv{k}_\para$, would then be determined only by quasiparticles near $\bm{k}_\parallel$.
When the $k$-space separation of the Fermi surfaces is larger than $\sim 1/\Delta r$, the interlayer connection would effectively vanish, and therefore the transmission of the supercurrent would become negligible.}

{
Figure \ref{fig:jcjd-theta} demonstrates that the Josephson critical current $J_\mathrm{c}$ can be effectively tuned by varying the twist angle, from a high-transparency JJ ($J_\mathrm{c} \approx J_\mathrm{d}$) in the small-angle regime to a low-transparency junction ($J_\mathrm{c} \ll J_\mathrm{d}$) in the large-angle regime.
This tunability of the coupling strength by twist angle could be highly beneficial for the design of Josephson devices.
}


\begin{figure}
\begin{center}
   \includegraphics [width=\linewidth]{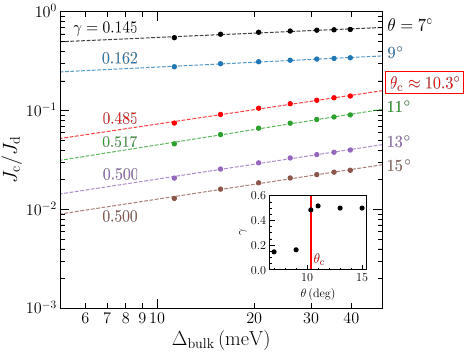}
   \caption{
   $J_\mathrm{c} / J_\mathrm{d}$ as a function of the bulk superconducting gap $\Delta_\mathrm{bulk}$ for several twist angles. 
   The dashed lines are fitting power functions, \rev{and the values of $\gamma$ are their exponents.
   The twist-angle dependence of the exponent $\gamma$ is illustrated in the inset.}
            }
            \label{fig:jcjd-delta}
 \end{center}
 \end{figure}

{
While we adopted the bulk superconducting gap $\Delta_\mathrm{bulk} \approx 20 \, \mathrm{meV}$ in the calculations above,  
it is important to examine the dependence of the supercurrent on the gap size, in view of realistic superconductors where $\Delta_\mathrm{bulk}$ is typically on the order of several meV.  
Figure~\ref{fig:jcjd-delta} presents a log-log plot of $J_\mathrm{c} / J_\mathrm{d}$ as a function of $\Delta_\mathrm{bulk}$,  
for several twist angles ranging from $\theta = 7^\circ$ to $15^\circ$.
{When we simply assume a power-law dependence, its exponent $\gamma$ is less than 0.2 when the Fermi surfaces are connected ($\theta < \theta_\mathrm{c}$), while it increases to approximately 0.5 after the Fermi surface overlap is lost ($\theta > \theta_\mathrm{c}$),  
corresponding to $J_\mathrm{c} / J_\mathrm{d} \propto \sqrt{\Delta_\mathrm{bulk}}$.}
\rev{At the critical twist angle ($\theta = \theta_\mathrm{c}$), the shape of the $J_\mathrm{c}/J_\mathrm{d}$ plot is similar to that at larger twist angles, $\theta = 11^\circ$, $13^\circ$, and $15^\circ$. 
This is because, at $\theta=\theta_\mathrm{c}$, the Fermi surfaces touch at a point and the perpendicular electronic transport is dominated by the non-local $(-\vv{k}_\para, \vv{k}_\para) \to (-\vv{k}'_\para, \vv{k}'_\para)$ transition. }
These results indicate that the finite supercurrent in the absence of Fermi-surface overlap remains appreciable for realistic bulk gap values on the order of a few meV.
}

\section{$\Gamma$-centered Fermi surface}
\label{sec:gamma}

\begin{figure}
\begin{center}
   \includegraphics [width=0.9\linewidth]{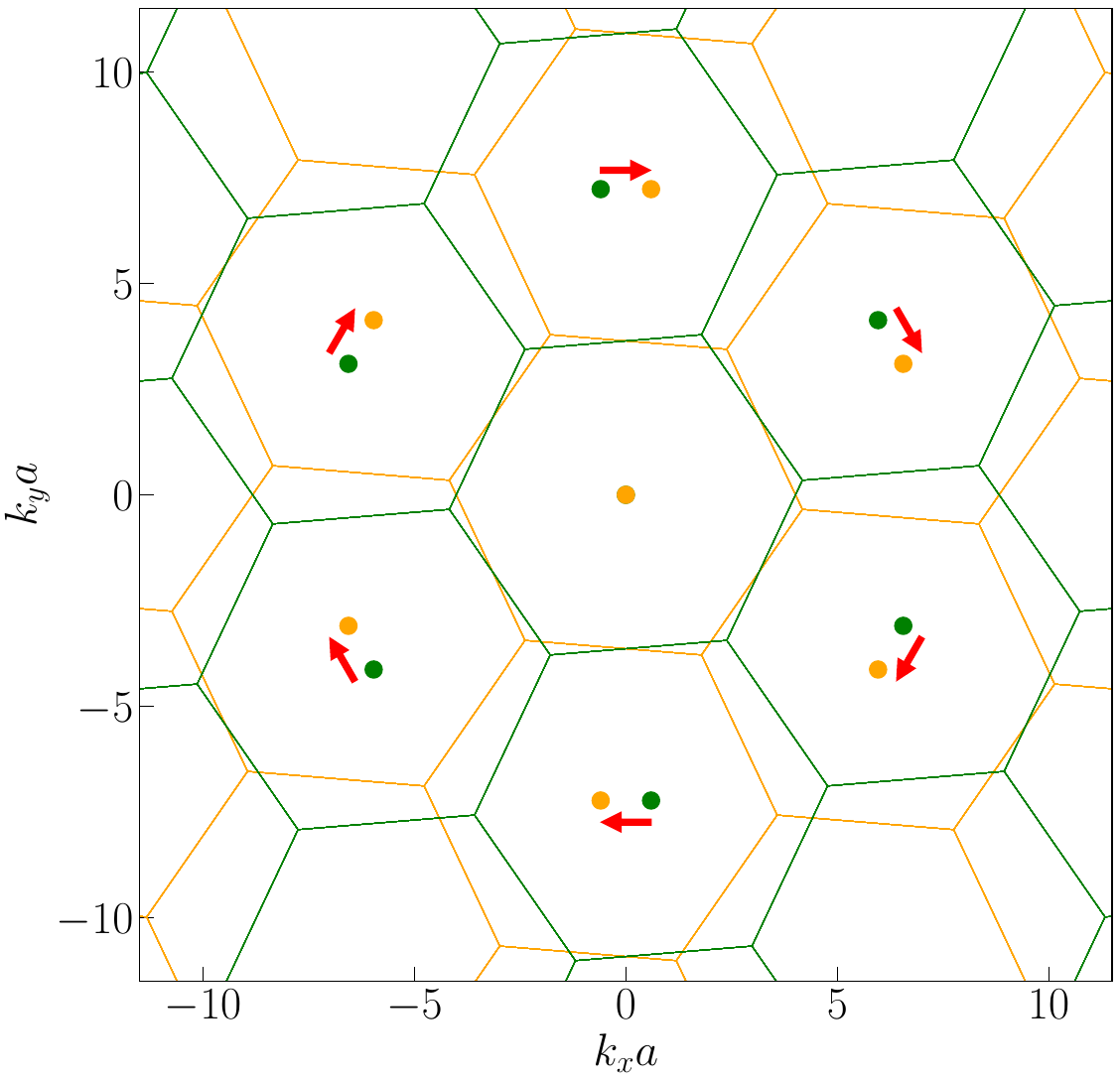}
   \caption{
   $\Gamma$ points with the Brillouin zones of the lower (yellow) and upper (green) honeycomb lattices in the extended-zone scheme.
   The central points are the completely overlapping $\Gamma$ points of each layer, giving the most dominant contribution \rev{to} the interlayer coupling.
   The red arrows represent the shift vectors between the two $\Gamma$ points being distant from the origin, whose contributions are negligibly small (see the text).
            }
            \label{fig:moirebz-gamma}
 \end{center}
 \end{figure}

{
If the Fermi surface of the bulk system is centered around the $\Gamma$ point, a twist does not lead to a separation of the Fermi surfaces between the lower and upper parts.  
In this section, we discuss the Josephson effect in such a case, i.e., a twisted interface between 3D superconductors with a $\Gamma$-centered Fermi surface.
Figure~\ref{fig:moirebz-gamma} shows the $\Gamma$ points of the upper and lower bulk parts in the extended zone scheme, where the central $\Gamma$ points remain aligned under twist~\cite{Fujimoto2022}.  
This is in sharp contrast to the $K$-pocket case (Fig.~\ref{fig:moirebz}), where the $K$ points are shifted by $\vv{q}_1$, $\vv{q}_2$, and $\vv{q}_3$.
}

{
In general, the interlayer coupling matrix element at the twisted interface takes the form $t(\vv{Q}) e^{i \bm{q} \cdot \bm{r}}$, where $\bm{q}$ is the Fermi surface shift vector, $\vv{Q}$ is the momentum of the Fermi surface in the extended Brillouin zone, and $t(\vv{k})$ is the Fourier transform of the real-space interlayer hopping amplitude~\cite{tani2023}.  
In the $K$-centered case, $\bm{q}$ corresponds to $\vv{q}_1$, $\vv{q}_2$, and $\vv{q}_3$, and $\vv{Q}$ is the $\vv{K}_\xi$ point, resulting in the interlayer coupling matrix elements in Eq.~\eqref{eq:interlayer}, where $t(K)$ corresponds to the parameters $u$ and $u'$.
In the $\Gamma$-pocket case, by contrast, the dominant contribution comes from $\vv{Q} = 0$ (the central points in Fig.~\ref{fig:moirebz-gamma}), where the shift $\bm{q}$ vanishes.  
This yields a uniform interlayer coupling $t(0)$, independent of the twist angle $\theta$.  
There are also correction terms arising from off-center points
in Fig.~\ref{fig:moirebz-gamma} with finite shift vectors $\bm{q}$ (indicated by red arrows), but their amplitudes $t(\vv{Q})$ are much smaller than the dominant contribution $t(0)$ because of large $\bm{Q}$~\cite{tani2023}.
From the above considerations, we expect nearly perfect electronic transmission across the interface in this case, and the system does not behave as a Josephson junction.
}

{
In realistic layered superconductors such as NbSe$_2$, Fermi surfaces around both the $\Gamma$ and $K$ points coexist~\cite{Doran1978,Lebegue2009,He2018,Habara2021}.
The effect of an interlayer twist on such a multipocket system can be discussed as follows.
When the system is viewed as an effective one-dimensional model with a fixed in-plane momentum $\vv{k}_\parallel$, the effective interface hopping parameter differs between the $\Gamma$ and $K$ pockets.
Let us define the interlayer hopping in the nontwisted and twisted bilayer units at the $\Gamma$ pocket as $t_0^{(\Gamma)}$ and $t_\mathrm{tw}^{(\Gamma)}$, respectively,
and those at the $K$ pocket as $t_0^{(K)}$ and $t_\mathrm{tw}^{(K)}$.
We denote the superconducting phase difference between adjacent layers (or bilayer units) in the nontwisted and twisted regions as $\Delta \varphi_0$ and $\Delta \varphi_\mathrm{tw}$, respectively.
These phase differences are assumed to be common to both the $\Gamma$ and $K$ channels.
As discussed above, the hopping amplitude in the $\Gamma$-channel is almost unaffected by a twist, and therefore we assume $t_\mathrm{tw}^{(\Gamma)} = t_0^{(\Gamma)}$.
Modeling the supercurrent in each channel as
\begin{equation}
\begin{split}
    & J^{(\Gamma)} = t_0^{(\Gamma)} \Delta \varphi_0,
    \\
    & J^{(K)} = t_0^{(K)} \Delta \varphi_0,
\end{split}
\end{equation}
the conservation of the total current $J^{(\Gamma)} + J^{(K)}$  in the perpendicular direction requires
\begin{equation}
(t_0^{(\Gamma)} + t_0^{(K)}) \Delta \varphi_0 = (t_0^{(\Gamma)} + t_\mathrm{tw}^{(K)}) \Delta \varphi_\mathrm{tw},
\end{equation}
which leads to
\begin{equation}
    \frac{\Delta\varphi_\mathrm{tw}}{\Delta\varphi_0} 
    =
    \frac{t_0^{(\Gamma)} + t_0^{(K)}}{t_0^{(\Gamma)} + t_\mathrm{tw}^{(K)}}.
    \label{eq:gamma-k-phase}
\end{equation}
}

{
Since the twist generally weakens the effective interlayer hopping [i.e., $t_\mathrm{tw}^{(K)} < t_0^{(K)}$], Eq.~\eqref{eq:gamma-k-phase} indicates that the interfacial phase difference $\Delta\varphi_\mathrm{tw}$ becomes larger than the bulk value $\Delta\varphi_0$.
In the presence of a $\Gamma$ pocket, this phase jump becomes milder than that in the case with only a $K$ pocket. Regarding the critical current, the depairing current of the bulk system is given by $J_\mathrm{d} = J_\mathrm{d}^{(\Gamma)} + J_\mathrm{d}^{(K)}$. For the twisted interface, we have $J_\mathrm{c} = J_\mathrm{d}^{(\Gamma)} + J_\mathrm{c}^{(K)}$, assuming that $J_\mathrm{c}^{(\Gamma)} = J_\mathrm{d}^{(\Gamma)}$ since the $\Gamma$ channel remains unaffected by the twist. Therefore, the ratio of the critical current to the depairing current becomes
\begin{equation}
\frac{J_\mathrm{c}}{J_\mathrm{d}} \approx
\frac{J_\mathrm{d}^{(\Gamma)} + J_\mathrm{c}^{(K)}}{J_\mathrm{d}^{(\Gamma)} + J_\mathrm{d}^{(K)}}.
\end{equation}
This indicates that the presence of a $\Gamma$ pocket enhances the ratio $J_\mathrm{c} / J_\mathrm{d}$. In previous experiments on NbSe$_2$~\cite{Yabuki2016}, it was reported that $J_\mathrm{c} / J_\mathrm{d} > 0.5$, a relatively large value that can be attributed to the substantial overlap of the $\Gamma$-pocket Fermi surfaces across the twisted interface.
}

{
If the $K$-pocket Fermi surfaces are completely separated by a sufficiently large twist, then the critical current is given by $J_\mathrm{c} = J_\mathrm{c}^{(\Gamma)} + J_\mathrm{c}^{(K)} \approx J_\mathrm{c}^{(\Gamma)}$.
In this case, we obtain
\begin{equation}
\frac{J_\mathrm{c}}{J_\mathrm{d}} \approx
\frac{J_\mathrm{d}^{(\Gamma)}}{J_\mathrm{d}^{(\Gamma)} + J_\mathrm{d}^{(K)}},
\end{equation}
which implies that observing $J_\mathrm{c} / J_\mathrm{d}$ in a twisted system allows one to extract the relative depairing current contributions from the two pockets. This serves as a kind of momentum-resolved probe of the superconducting state.
}

\section{Conclusions}
\label{sec:conclusion}

{
We have developed a theoretical formulation for the BdG equation, the gap equation, and the perpendicular supercurrent in twisted 3D superconductors.
To focus on the effects of twisting on superconductivity, our method was applied to a toy model Hamiltonian whose Fermi surfaces are located around the BZ corner.
By calculating the superconducting phases, we confirm the presence of a phase jump at the twisted interface, thereby realizing a twist JJ.
The twist-angle dependence of the perpendicular supercurrent reveals the high tunability of the Josephson critical current via twisting.
It is further elucidated that finite interface coupling and a nonzero supercurrent persist even when the Fermi surfaces are separated by a twist.
This behavior is completely different from that observed in twisted 3D metals, where the normal perpendicular conduction vanishes upon Fermi-surface separation.

We also examined how $J_\mathrm{c}/J_\mathrm{d}$ depends on the bulk superconducting gap $\Delta_\mathrm{bulk}$.
The result implies that a finite supercurrent can persist even without Fermi-surface overlap, and is expected to remain sizable for realistic gap magnitudes on the order of a few meV.
}

{
Lastly, we analyzed a twisted junction in which the Fermi surface is centered at the Brillouin zone $\Gamma$ point. We found that the twist-induced modulation of the interlayer coupling is minimal in this case, leading to little suppression of the Josephson current.
In systems like NbSe$_2$ that host both $\Gamma$- and $K$-pocket Fermi surfaces, we showed that the contribution from the $\Gamma$ pocket significantly enhances $J_\mathrm{c}/J_\mathrm{d}$, thereby highlighting the importance of the Fermi-surface geometry in determining the junction properties.
}

{
Our formulation can be readily extended to a wide range of twisted 3D superconducting junctions, including those with barrier layers inserted between the two superconductors. Notably, introducing a ferromagnetic barrier may provide a unique opportunity to study how the Josephson diode effect emerges or is modified by twisting.
}

\begin{acknowledgements}
This work was supported by Japan JSPS KAKENHI Grants No.~JP23KJ1497, No.~JP20K14415, No.~JP24K06921, No.~JP20H01840, No.~JP20H00127, No.~JP21H05236, No.~JP21H05232, No.~JP25K00938 and by JST CREST Grant No.~JPMJCR20T3. 
\end{acknowledgements}

\appendix
\section{Continuum model for BdG equation}
\label{app:valley}
{In this appendix, we describe the continuum model that underlies the Bogoliubov–de Gennes (BdG) formalism used in the main text. We begin with the interacting tight-binding Hamiltonian 
\begin{equation}
\hat{H} = \hat{H}_0 + \hat{H}_\mathrm{int},
\end{equation}
where the first term,
\begin{equation}
    \hat{H}_0 = \sum_{s=\uparrow,\downarrow}\sum_{\bm{r},\bm{r}'} \hat{\psi}^\dagger_{s}(\vv{r}) H_0(\vv{r}, \vv{r}') \hat{\psi}_{s}(\vv{r}'),
\end{equation}
represents the single-particle part of the Hamiltonian, 
describing electron hopping from site $\vv{r}'$ to $\vv{r}$.
The second term, 
\begin{equation}
    \hat{H}_\mathrm{int} = -V \sum_{\vv{r}} \hat{\psi}^\dagger_{\uparrow}(\vv{r}) \hat{\psi}^\dagger_{\downarrow}(\vv{r}) \hat{\psi}_{\downarrow}(\vv{r}) \hat{\psi}_{\uparrow}(\vv{r}),
\end{equation}
describes an on-site attractive interaction between electrons, where $V>0$ is the interaction strength.
The field operator $\hat{\psi}_{s}(\vv{r})$ annihilates an electron with spin $s=\uparrow, \downarrow$ at site $\vv{r}$.}

{
To describe superconductivity, we apply the mean-field approximation with an on-site $s$-wave pairing assumption. 
The corresponding superconducting order parameter is defined as
\begin{equation}\label{eq:delappendix}
    \Delta(\vv{r}) = -V\langle \hat{\psi}_{\downarrow}(\vv{r}) \hat{\psi}_{\uparrow}(\vv{r}) \rangle.
\end{equation}
Under this approximation, the Hamiltonian reduces to a mean-field form
\begin{equation}
    \hat{H}\sim \hat{H}_\mathrm{MF}
    =
    \sum_{\vv{r}, \vv{r}'}
    \mqty(\hat{\psi}_{\uparrow}^\dagger(\vv{r}), & \hat{\psi}_{\downarrow}(\vv{r}))
    \, H^\mathrm{BdG} (\vv{r}, \vv{r}')
    \,
    \mqty(\hat{\psi}_{\uparrow}(\vv{r}') \\ 
    \hat{\psi}_{\downarrow}^{\dagger}(\vv{r}')), \label{eq:H-MF}
\end{equation}
with the corresponding BdG Hamiltonian
\begin{equation}\label{eq:bdgappendix}
H^\mathrm{BdG} (\vv{r}, \vv{r}') =
    \mqty(
    H_0(\vv{r}, \vv{r}') & 
    \Delta(\vv{r}) \delta_{\vv{r},\vv{r}'} \\
    \Delta^{*}(\vv{r}) \delta_{\vv{r},\vv{r}'} & 
    -H^{*}_0 (\vv{r}, \vv{r}')
    ),
\end{equation}
expressed in the Nambu spinor basis.
}

{
The BdG equation in this tight-binding form is given by 
\begin{equation}
\label{eq:tb-bdgeq}
    \sum_{\vv{r}'}
    H^\mathrm{BdG} (\vv{r}, \vv{r}')
    \, \mqty(u^\nu(\vv{r}') \\  v^\nu(\vv{r}')) 
    = E^\nu \, \mqty(u^\nu(\vv{r}) \\  v^\nu(\vv{r})),
\end{equation}
where $( u^{\nu}, v^{\nu})^\top$ is the quasiparticle wavefunction labeled by a quantum number $\nu$. 
The superconducting order parameter in Eq.~\eqref{eq:delappendix} is determined self-consistently from the solution of the BdG equation as 
\begin{align}
\label{eq:tb-gapeq}
    \Delta(\vv{r}) = -V \sum_\nu 
    v^{\nu*} (\vv{r}) u^{\nu} (\vv{r}) f(E^\nu),
\end{align}
where $f(E)$ is the Fermi--Dirac distribution function.
}

{
To derive the continuum model, we consider a system with sublattice $X$ (equivalent to $X_n$ in the main text) located at $\bm{r}=\bm{r}_X=\bm{\tau}_X+\vv{R}$, where $\vv{R}=m_1 \vv{a}_1 + m_2 \vv{a}_2$ is a 2D lattice vector.
Assuming that the system possesses two valleys 
at Bloch wave vector $\vv{k}=\vv{K}_\xi$ with $\xi=\pm1$ and $\vv{K}_{-\xi}=-\vv{K}_{\xi}$,
the quasiparticle wavefunction can be written as
\begin{equation}
\label{eq:continuum-approx}
    \begin{split}
        &   u^\nu (\vv{r}_{X}) = \sum_{\xi=\pm} e^{i \vv{K}_\xi \vdot \vv{r}_X} 
        u_{X\xi}^\nu (\vv{r}_X),
        \\
        &   v^\nu (\vv{r}_{X}) = \sum_{\xi=\pm} e^{-i \vv{K}_{-\xi} \vdot \vv{r}_X} 
        v_{X\xi}^\nu (\vv{r}_X),
    \end{split}
\end{equation}
where $u^\nu_{X\xi} (\vv{r})$ and $v^\nu_{X\xi} (\vv{r})$ 
are slowly varying envelope function in real space. 
}


{
We assume the intervalley coupling in the normal-state Hamiltonian $H _0(\vv{r}, \vv{r}')$ is negligible, as in graphene. 
Furthermore, under the assumption of a vanishing Cooper-pair momentum, 
the superconducting order parameter $\Delta (\vv{r})=\Delta_{X}(\vv{r})$ couples electrons from opposite valleys. 
As a result, the BdG Hamiltonian~[Eq.~\eqref{eq:bdgappendix}] is block diagonalized for $\xi$, which thus serves as a good quantum number labeling the quasiparticle.
From Eqs.~\eqref{eq:tb-bdgeq} and \eqref{eq:continuum-approx}, the BdG equation becomes
\begin{equation}
\label{eq:bdgeq-continuous}
\begin{split}
    \sum_{X'}
    \mqty(
    H_{\xi XX'} (-i\nabla) & 
    \Delta_X(\vv{r}) \delta_{XX'} \\
    \Delta^{*}_X(\vv{r}) \delta_{XX'} & 
    -H_{-\xi XX'}^{*} (-i\nabla))
    \, \mqty(u^\nu_{X'\xi} (\vv{r}) \\  v^\nu_{X' \xi}(\vv{r})) &
    \\
    = E^\nu_\xi \, \mqty(u^\nu_{X\xi} (\vv{r}) \\  v^{\nu}_{X \xi} (\vv{r})) &,
\end{split}
\end{equation}
{where we introduce the Bloch Hamiltonian  
$H_{\xi XX'}(\vv{k})=\sum_{\vv{R}'}e^{-i\bm{\tau}_X \cdot (\bm{k}+\bm{K}_\xi)}H_0(\vv{\tau}_X,\vv{\tau}_{X'}+\vv{R}')e^{i(\bm{\tau}_{X'}+\vv{R}') \cdot (\bm{k}+\bm{K}_\xi)}$. Within the continuum approximation, we replace $\vv{k}\rightarrow-i\bm{\nabla}$ and treat $\vv{r}$ as a continuous variable.} 
Consequently, the self-consistent gap equation~[Eq.~\eqref{eq:tb-gapeq}] becomes
\begin{equation}
\label{eq:tb-gapeq-continuous}
    \Delta_X(\vv{r}) = 
    -V \sum_{\xi=\pm} \sum_\nu 
    v^{\nu*}_{X \xi} (\vv{r}) u_{X\xi}^{\nu} (\vv{r}) f(E^\nu_\xi).
\end{equation}
Bloch theorem 
\begin{equation}
\label{eq:bloch-thm}
    \begin{split}
        &   u_{X\xi}^{\nu} (\vv{r}) 
        = e^{i \vv{k} \vdot \vv{r}} u_{X\xi \vv{k}}^{\nu} (\vv{r})  
        = e^{i \vv{k} \vdot \vv{r}} u^\alpha_{X} (\vv{r}) ,
        \\
        &   v_{X\xi}^{\nu} (\vv{r}) 
        = e^{i \vv{k} \vdot \vv{r}} v_{X\xi \vv{k}}^{\nu} (\vv{r})  
        = e^{i \vv{k} \vdot \vv{r}} v^\alpha_{X} (\vv{r}) ,
        \\
    \end{split}
\end{equation}
leads to Eqs.~\eqref{eq:twist-bdg-eq} and \eqref{eq:gapeq-realspace} in the main text.
Here, $\alpha$ is a joint index for $(\nu, \xi, \vv{k})$.
}

\section{Continuum model for supercurrent}
\label{app:supercurrent}
Here, we illustrate how the perpendicular supercurrent is formulated in the continuum model. First, let us consider the Heisenberg equation within a general tight-binding model system described in Appendix~\ref{app:valley},
\begin{equation}
\label{eq:heisenberg}
\begin{split}
    & \dv{t} \hat{\rho}_n (\vv{r}) = \frac{i}{\hbar}  \comm{\hat{H}_\mathrm{MF}}{\hat{\rho}_n (\vv{r})},
    \\
    & \hat{\rho}_n (\vv{r}) = -e \sum_{X,s} 
    \hat{\psi}_{X_n s}^\dagger (\vv{r})
    \hat{\psi}_{X_n s} (\vv{r}),
\end{split}
\end{equation}
where $\hat{\rho}_n (\vv{r})$ denotes the electron density operator at a position $\vv{r}$ of a bilayer unit $n$.
We define the perpendicular current density operator $\hat{J}_{n+1, n} (\vv{r}, \vv{r}')$ through the equation of continuity,
\begin{equation}
\label{eq:current-operator-def}
    \dv{t} \hat{\rho}_n (\vv{r}') = - \sum_{\vv{r}}
    \frac{\hat{J}_{n+1, n} (\vv{r}, \vv{r}') - \hat{J}_{n, n-1}(\vv{r}', \vv{r})}{c},
\end{equation}
Equations~\eqref{eq:heisenberg} and \eqref{eq:current-operator-def} lead to
\begin{widetext}
\begin{equation}
\begin{split}
    \hat{J}_{n+1, n} (\vv{r}, \vv{r}') 
    &= \frac{iec}{\hbar} \sum_{X \tilde{X} s}
    \left[
    H_{0 X_{n+1} \tilde{X}_n} (\vv{r}, \vv{r}')
    \hat{\psi}_{X_{n+1}s}^{\dagger} (\vv{r})
    \hat{\psi}_{\tilde{X}_{n}s} (\vv{r}') 
    -
    H_{0 X_{n} \tilde{X}_{n+1}} (\vv{r}', \vv{r})
    \hat{\psi}_{X_{n}s}^{\dagger} (\vv{r}')
    \hat{\psi}_{\tilde{X}_{n+1}s} (\vv{r}) 
    \right]
    \\
    & \quad -\frac{2iec}{\hbar} \sum_{n' = n_0}^{n} \sum_X \delta_{\vv{r}, \vv{r}'}
    \left[
    \Delta_{X_{n'}}(\vv{r}) 
    \hat{\psi}_{X_{n'} \uparrow}^\dagger(\vv{r}) 
    \hat{\psi}_{X_{n'} \downarrow}^\dagger(\vv{r}) 
    - \Delta_{X_{n'}}^{*} (\vv{r})
    \hat{\psi}_{X_{n'} \downarrow}(\vv{r}) 
    \hat{\psi}_{X_{n'} \uparrow}(\vv{r}) 
    \right],
\end{split}
\end{equation}
where $n_0$ is an arbitrary reference bilayer unit.
The expectation value of the current density, $J_n (\vv{r}, \vv{r}') \equiv \expval{\hat{J}_{n+1, n} (\vv{r}, \vv{r}')}$, is then given by
\begin{equation}
\begin{split}
    J_n (\vv{r}, \vv{r}') 
    &= 
    \frac{-2ec}{\hbar} \Im \sum_{X \tilde{X} \nu}
    H_{0 X_{n+1} \tilde{X}_n} (\vv{r}, \vv{r}')
    \left[
    u_{X_{n+1}}^{\nu*} (\vv{r})
    u_{\tilde{X}_{n}}^{\nu} (\vv{r}') f(E^\nu)
    +
    v_{X_{n+1}}^{\nu} (\vv{r})
    v_{\tilde{X}_{n}}^{\nu*} (\vv{r}') f(-E^\nu)
    \right]
    \\
    & \quad +\frac{4ec}{\hbar} \Im \sum_{n' = n_0}^{n} \sum_X
    \delta_{ij} \Delta_{X_{n'}}(\vv{r})
    \left[
    \sum_\nu u_{X_{n'}}^{\nu*} (\vv{r}) v_{X_{n'}}^{\nu}(\vv{r})
    f(E^\nu) 
    \right].
\end{split}
\end{equation}
By the continuum approximation described in Appendix~\ref{app:valley}, we have
\begin{equation}
\label{eq:current-continuous}
\begin{split}
    J_n(\vv{r}) =& \frac{-2ec}{\hbar} \Im
    \sum_{\xi=\pm} \sum_{\nu} \sum_{X, \tilde{X}}
    \left[
    u_{X_{n+1} \xi}^{\nu*}(\vv{r}) 
    H_{\xi X_{n+1} \tilde{X}_n} (-i \nabla)
    u_{\tilde{X}_{n} \xi}^{\nu}(\vv{r}) 
    f(E^{\nu}_{\xi})
    + 
    v_{X_{n+1} \xi}^{\nu} (\vv{r}) 
    H_{\xi X_{n+1} \tilde{X}_n} (-i \nabla)
    v_{\tilde{X}_{n} \xi}^{\nu*} (\vv{r}) 
    f(-E^{\nu}_{\xi})
    \right]
    \\
    &   + \frac{4ec}{\hbar}
    \left( -\frac{1}{V} \right)
    \Im
    \sum_{n' = n_0}^{n} \sum_{X}
    \Delta_{X_{n'}} (\vv{r})
    \left[ -V
    \sum_{\xi=\pm}
    \sum_{\nu} 
    v_{X_{n'} \xi}^{\nu*}(\vv{r}) 
    u_{X_{n'} \xi}^{\nu}(\vv{r}) 
    f(E^{\nu}_{\xi})
    \right]^{*}.
\end{split}
\end{equation}
\end{widetext}
In general, the second term known as a source term is necessary for the current conservation~\cite{furusaki1991dc,furusaki1994dc,Asano2001numerical}.
In a self-consistent loop of the numerical calculation, we first compute wavefunctions $(u(\vv{r}), v(\vv{r}))$ from the BdG equation [Eq.~\eqref{eq:bdgeq-continuous}] using an old order parameter $\Delta (\vv{r})$.
Subsequently, from the gap equation [Eq.~\eqref{eq:tb-gapeq-continuous}], a new order parameter is obtained; this loop is iterated until the old and new order parameters match.
Given this procedure, in the second term of Eq.~\eqref{eq:current-continuous}, the new order parameter $[\cdots]^*$ is equal to the old $\Delta_{X_{n'}}^* (\vv{r})$ only when the self-consistent states are used, giving the vanishment of the second term [due to $\Im |\Delta (\vv{r})|^2 = 0$].

The Bloch's theorem [Eq.~\eqref{eq:bloch-thm}] leads to the expression of $J_n(\vv{r}_\para)$ in the main text [Eq.~\eqref{eq:current-realspace}] where the source term is absent, as we use the self-consistent solution throughout the paper.
This current density exhibits a moir\'e-scale modulation and can therefore be expanded in terms of the wavevector $\vv{q}$,
\begin{equation}
    J_n(\vv{r}_\para) = \sum_{\vv{q}} 
    J_{n\vv{q}} e^{i \vv{q} \vdot \vv{r}_\para}.
\end{equation}
By substituting Eq.~\eqref{eq:fourier-wavefunc} into Eq.~\eqref{eq:current-realspace}, the Fourier component $J_{n \vv{q}}$ is obtained as
\begin{equation}
    \begin{split}
        &   J_{n\vv{q}} = \frac{1}{2}
        \left[
        \Im(L_{n\vv{q}} + L_{n, -\vv{q}}) -i \Re(L_{n\vv{q}} - L_{n, -\vv{q}})
        \right],
        \\
        &   L_{n\vv{q}} = \frac{-2e}{\hbar}
        \sum_{\alpha} \sum_{X, \tilde{X}}
        \sum_{\vv{q}', \vv{q}''}
        \\
        &   \quad \left\{
        u_{X_{n+1} \vv{q}''}^{\alpha*} 
        [\mathcal{H}_{\xi, \vv{q}-\vv{q}'+\vv{q}'' } (\vv{k}_\para)]_{X_{n+1}, \tilde{X}_n}
        u_{\tilde{X}_{n} \vv{q}'}^{\alpha} 
        f(E^\alpha)
        + \right.
        \\
        &   \quad\left. v_{X_{n+1} \vv{q}'}^{\alpha} 
        [\mathcal{H}_{\xi, \vv{q}-\vv{q}'+\vv{q}'' } (-\vv{k}_\para)]_{X_{n+1}, \tilde{X}_n}
        v_{\tilde{X}_{n} \vv{q}''}^{\alpha*} 
        f(-E^\alpha)
        \right\},
    \end{split}
\end{equation}
where $\mathcal{H}_{\xi \vv{q}} (\vv{k}_\para)$ is the Fourier component of the normal-state Hamiltonian [Eq.~\eqref{eq:twist-normal-hamil}], defined as
\begin{equation}
    \mathcal{H}_{\xi} (\vv{k}_\para) =
    \sum_{\vv{q}}
    \mathcal{H}_{\xi \vv{q}} (\vv{k}_\para)
    e^{i \vv{q} \vdot \vv{r}_\para}.
\end{equation}
For example, the $4\times4$ block $[\mathcal{H}_{\xi \vv{q}} (\vv{k}_\para)]_{X_{N+1}, \tilde{X}_{N}} = T_\mathrm{int}$ [Eq.~\eqref{eq:interlayer}] contains non-zero components for $\vv{q} = \vv{q}_1, \vv{q}_2, \vv{q}_3$.

\section{Depairing current}
\label{app:depairing}

The depairing current is obtained as the bulk Josephson critical current.
To perform the calculation, we set the boundary condition for the 3D Hamiltonian [Eq.~\eqref{eq:3d-bdg}] as follows.
We solve the $8\times8$ BdG equation
\begin{equation}
\label{eq:3d-bdgeq}
    \mqty(
    H^\mathrm{3D}_\xi (\vv{k}_\para, k_z) & \underline{\Delta} \\
    \underline{\Delta}^* & -H^{\mathrm{3D}*}_{-\xi} (-\vv{k}_\para, -k_z)
    )
    \,
    \mqty(\vv{u}_{k_z}^{\alpha} \\ \vv{v}_{k_z}^{\alpha}) 
    = E_{k_z}^{\alpha} \,
    \mqty(\vv{u}_{k_z}^{\alpha} \\ \vv{v}_{k_z}^{\alpha}),
\end{equation}
where $\vv{u}_{k_z}^{\alpha} = (u_{A k_z}^{\alpha}, u_{B k_z}^{\alpha}, u_{A' k_z}^{\alpha}, u_{B' k_z}^{\alpha})^\top$; $\vv{v}^\alpha_{k_z}$ is defined in the same manner.
To introduce a phase shift $\delta\varphi_\mathrm{bulk}$ between adjacent bilayer units, we replace matrix elements as
\begin{equation}
    T^\dagger e^{i k_z c} \rightarrow T^\dagger e^{i k_z c + i\delta\varphi_\mathrm{bulk}}
\end{equation} 
for the electron part $H^\mathrm{3D}_\xi (\vv{k}_\para, k_z)$, and
\begin{equation}
    -(T^\dagger)^{*} e^{i k_z c} \rightarrow -(T^\dagger)^{*} e^{i k_z c - i\delta\varphi_\mathrm{bulk}}
\end{equation}
for the hole part $-H^{\mathrm{3D}*}_{-\xi} (-\vv{k}_\para, -k_z)$.

Under this boundary condition, we perform the self-consistent calculation of the BdG equation [Eq.~\eqref{eq:3d-bdgeq}] and the gap equation
\begin{equation}
    \Delta_{X} =
    -V \sum_\alpha 
    \int
    \frac{\dd k_z}{2\pi} \,
    v_{X k_z}^{\alpha*}
    u_{X k_z}^{\alpha}
    f(E^{\alpha}_{k_z}).
\end{equation}
From the obtained solution, the perpendicular supercurrent $J_\mathrm{bulk}$ is calculated from
\begin{equation}
\begin{split}
    &   J_\mathrm{bulk} = \frac{-2e}{\hbar} \Im
    \sum_{\alpha} \sum_{X, \tilde{X}}
    \int \frac{\dd k_z}{2\pi}
    \\
    &   \left\{
    u_{X k_z}^{\alpha*} e^{ik_zc + i\delta\varphi_\mathrm{bulk}}
    [
    \mathcal{H}^\mathrm{3D}_\xi(\vv{k}_\para, k_z)]_{X \tilde{X}}
    u_{\tilde{X} k_z}^{\alpha} 
    f(E^{\alpha}_{k_z}) \right.
    \\
    &   \left. + 
    v_{X k_z}^{\alpha}  e^{ik_zc - i\delta\varphi_\mathrm{bulk}}
    [\mathcal{H}^\mathrm{3D}_\xi(-\vv{k}_\para, -k_z)]_{X \tilde{X}}
    v_{\tilde{X} k_z}^{\alpha*} 
    f(-E^{\alpha}_{k_z})
    \right\}.
\end{split}
\end{equation}
In the small $\delta\varphi_\mathrm{bulk}$ regime, the supercurrent $J_\mathrm{bulk}$ increases as $\delta\varphi_\mathrm{bulk}$ is enhanced, while the superconducting gap $|\Delta_X|$ remains unchanged.
As $\delta\varphi_\mathrm{bulk}$ is more increased, $J_\mathrm{bulk}$ reaches a peak, and subsequently $J_\mathrm{bulk}$ and $|\Delta_X|$ rapidly drop. 
This behavior is attributed to an instability of a finite-momentum Cooper pair, which is depaired for a too large $\delta\varphi_\mathrm{bulk}$.
Therefore, the maximum value of the current-phase relation $J_\mathrm{bulk}(\delta\varphi_\mathrm{bulk})$ is identified with the depairing current $J_\mathrm{d}$~[Eq.~\eqref{eq:jd-definition}].

\section{Alternative models}
\label{app:alternative}

\subsection{Massive model}

\begin{figure*}
\begin{center}
   \includegraphics [width=0.7\linewidth]{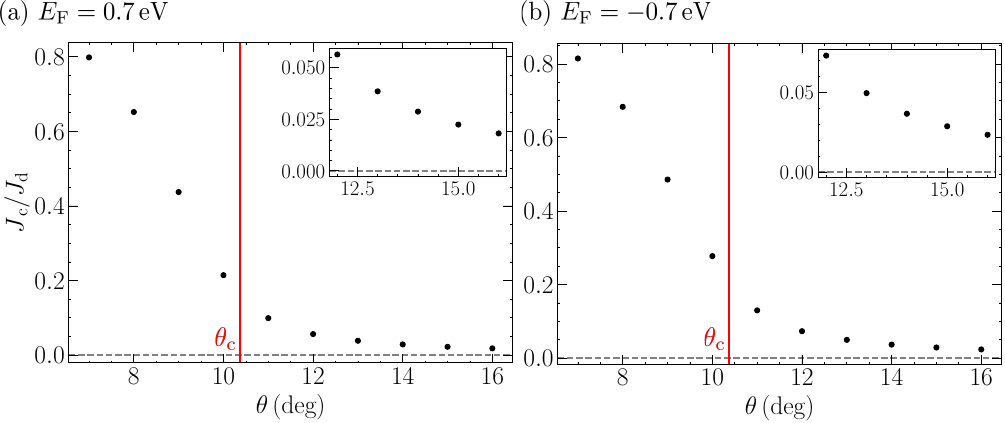}
   \caption{
   $J_\mathrm{c} / J_\mathrm{d}$ ratio in the massive model for (a) $E_\mathrm{F} = 0.7 \, \mathrm{eV}$ and (b) $E_\mathrm{F} = -0.7 \, \mathrm{eV}$.
   The vertical red lines indicate the critical angle $\theta_\mathrm{c}$.
               }
            \label{fig:jcjd-theta-massive}
 \end{center}
 \end{figure*}

In the main text, we used Eq.~\eqref{eq:3d-normal-hamil} as a model Hamiltonian.
Here, we add an extra mass term with $m = 0.5 \, \mathrm{eV}$ and show this does not influence the qualitative result.
Specifically, the onsite energy $m$ ($-m$) is added to the $A$ and $A'$ ($B$ and $B'$) sublattices.
The mass term gives a sublattice unbalance, and therefore breaks a topological pseudospin texture with a finite winding number around $K_\xi$ point.

We investigate two models: an electron-doped model $E_\mathrm{F} = 0.7 \, \mathrm{eV}$ and a hole-doped model $E_\mathrm{F} = -0.7 \, \mathrm{eV}$.
In the former (latter), the Fermi surface consists of the conduction (valence) bands.
We use the electron-electron interaction strength of $V = 0.45 \, \mathrm{eV}$ to obtain $\Delta_A \approx 20 \, \mathrm{meV}$ for the model with $E_\mathrm{F} = 0.7 \, \mathrm{eV}$ ($\Delta_{B'} \approx 20 \, \mathrm{meV}$ for the model with $E_\mathrm{F} = -0.7 \, \mathrm{eV}$).
Figure~\ref{fig:jcjd-theta-massive} shows the calculated $J_\mathrm{c} / J_\mathrm{d}$ ratio as a function of the twist angle $\theta$.
We confirm a similar behavior as the massless model in the main text, where the finite supercurrent resides for $\theta > \theta_\mathrm{c}$.


\subsection{AA-stacking model}
Subsequently, we examine an AA-stacking model.
We employ a model Hamiltonian,
\begin{align}
    \begin{split}
        H^\mathrm{3D}_\xi (\vv{k}_\para, k_z) &= H_\xi(\vv{k}_\para) + T e^{-i k_zc/2} + T^\dagger e^{i k_zc/2} - E_\mathrm{F},
        \\
        H_\xi (\vv{k}_\para) &=
        \mqty(m & -v k_{-} \\
        -v k_{+} & -m
        ),  \quad
        T = 
        \mqty(t & 0 \\
        0 & t
        ),
    \end{split}
\end{align}
where the size of the matrices are reduced to $2\times2$ since a single layer forms a unit.
Along out-of-plane wavenumber $k_z$, the energy band has the dispersion of $2t \cos{(k_z c/2)}$.
This leads to a fairly large radius of the Fermi surface in $k_x k_y$ plane, and therefore the lower and upper Fermi surfaces are not separated by a twist of $\theta \approx 10^\circ$.
The Fermi-surface disconnection is realized by more shallow doping (smaller $|E_\mathrm{F}|$), however, such a Fermi energy traverses both the conduction and valence bands.
Since such a complex electronic structure is out of our interest, we modify the interlayer hopping parameters $t, u, u'$ to $1/4$ of the original values. 
Namely, we employ $t = 0.1 \, \mathrm{eV}$, $u = 0.0199 \, \mathrm{eV}$, and $u' = 0.0244 \, \mathrm{eV}$.
Using these parameters, this AA-stacking model becomes similar to the AB-stacking massless/massive models, where the critical twist angle becomes $\theta_\mathrm{c} \approx 10.35^\circ$ with $E_\mathrm{F} = -0.75 \, \mathrm{eV}$.

\begin{figure*}
\begin{center}
   \includegraphics [width=0.45\linewidth]{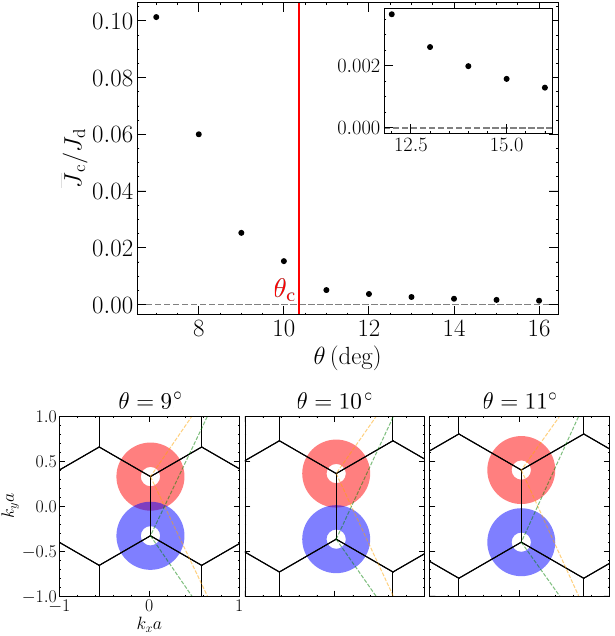}
   \caption{
   (Top)~$J_\mathrm{c} / J_\mathrm{d}$ in the AA-stacking massive model.
   The vertical red line represents the critical angle $\theta_\mathrm{c}$.
   (Bottom)~Fermi surfaces for several twist angles, projected onto the $k_x k_y$ plane.
   The red (blue) ring represents the Fermi surface of the lower (upper) part of the system.
               }
            \label{fig:jcjd-theta-AA}
 \end{center}
 \end{figure*}

In the top panel of Fig.~\ref{fig:jcjd-theta-AA}, $J_\mathrm{c} / J_\mathrm{d}$ is illustrated as a function of the twist angle $\theta$.
The electron-electron interaction strength $V = 0.45 \, \mathrm{eV}$ is employed to obtain $\Delta_{B} \approx 20 \, \mathrm{meV}$.
We can confirm a similar behavior to the AB-stacking massless or massive models.
The bottom panel of Fig.~\ref{fig:jcjd-theta-AA} illustrates the cylindrical Fermi surfaces projected onto the $k_x k_y$ plane.

\bibliography{reference}
\end{document}